\documentclass[notitlepage,superscriptaddress, longbibliography]{revtex4-1}

\usepackage{natbib}
\usepackage{graphicx}
\usepackage{amssymb}
\usepackage{amsmath}
\usepackage{ifthen}
\usepackage{braket}
\usepackage{xcolor}
\usepackage{bm}
\usepackage{bbm}
\usepackage{bbm}
\usepackage{maybemath}
\usepackage{hyperref}
\usepackage{comment}
\usepackage{siunitx}
\usepackage{dcolumn}
\newcolumntype{d}[1]{D{.}{.}{#1}}

\definecolor{garrosgreen}{rgb}{0.1, 0.4, 0.1}
\definecolor{dartmouthgreen}{rgb}{0.05, 0.5, 0.06}
\definecolor{duelferred}{rgb}{0.7, 0.2, 0.1}
\definecolor{cambridgeblue}{rgb}{0.1, 0.3, 1.0}
\definecolor{oxfordblue}{rgb}{0.05, 0.2, 0.7}

\def\calN{{\mathcal{N}}}
\def\calG{{\mathcal{G}}}
\def\calO{{\mathcal{O}}}

\newcommand{\ee}{{\mathrm e}}
\newcommand{\ii}{{\mathrm i}}
\newcommand{\dd}{{\mathrm d}}

\newcommand{\hate}{\hat{\mathrm{e}}}

\newcommand{\addrRolla}{Department of Physics and LAMOR,
Missouri University of Science and Technology,
Rolla, Missouri 65409, USA}

\newcommand{\addrDebrecen}{Hungarian Academy Institute
for Nuclear Physics (ATOMKI), Debrecen, Hungary}

\begin{document}

\title{Dispersion of Ultra--Relativistic Tardyonic and\\
Tachyonic Wave Packets on Cosmic Scales}

\author{Jos\'{e} Nicasio}
\affiliation{\addrRolla}

\author{Ulrich D.~Jentschura}
\affiliation{\addrRolla}
\affiliation{\addrDebrecen}

\begin{abstract}
We investigate the time propagation of
tachyonic (superluminal) and tardyonic (subluminal, ordinary)
massive wave packets on cosmic scales.
A normalizable wave packet cannot be monochromatic in 
momentum space and thus acquires 
a positional uncertainty (or packet width) that increases with
travel distance.
We investigate the question of how this 
positional uncertainty affects the uncertainty in the 
detection time for cosmic radiation on Earth.
In the ultrarelativistic limit, we find a 
unified result, $\delta x(t)/c^3 = m^2 \, \delta p \, t /p_0^3$,
where $\delta x(t)$ is the positional uncertainty,
$m$ is the mass parameter, $\delta p$ is the
initial momentum spread of the wave function,
and $p_0$ is the central momentum of the wave packet,
which, in the ultrarelativistic limit, is equal to its energy.
This result is valid for tachyons and tardyons;
its interpretation is being discussed.
\end{abstract}

\maketitle


%
%
\section{Introduction}
\label{sec1}

{\em Aficionados} of quantum field theory 
are interested in the dispersion of ultra-relativistic wave packets on cosmic
scales.  The unexpected early neutrino burst from the supernova 1987A as
reported in~\cite{DaEtAl1987} still inspires speculations about a potentially tachyonic
nature of at least some of the known neutrino 
species~\cite{Eh2012}. While this possibility
needs to be considered as remote, 
some interesting conclusions regarding fundamental properties 
of neutrinos have been derived from the observations~\cite{BaCo1987,ArRo1987,Lo1988}.
The situation gives rise to further
questions: Given the dispersion of quantum mechanical wave packets due to the
spreading of the momentum-space components, one might speculate that, over very
long (cosmic) distance and time scales, the quantum-mechanical wave packet
might have spreaded sufficiently to ``mimic early arrival'' of neutrino bursts,
due to position uncertainty after propagation over cosmic scales.

It is well known that, depending on the way in which 
one interprets the experiments, tiny violations of the causality principle
are permissible under the laws of quantum mechanics.
Let us consider a wave packet which fulfills the 
massless one-dimensional Klein--Gordon equation,
\begin{equation}
\label{start}
\phi(t, x) = \frac{1}{(2 \pi)^{1/4} \sqrt{ \delta x}} \, 
\exp\left[ -\frac{(x-ct)^2}{4 \delta x^2} \right] \, \cos(x-ct) \,,
\qquad
\left( \frac{1}{c^2} \frac{\partial^2}{\partial t^2} - 
\frac{\partial^2}{\partial x^2} \right) \, \phi(t, x) = 0 \,,
\end{equation}
which is nonmalized to the condition $\int \dd x \, | \phi(t=0, x) |^2 
= \int \dd x \, | \phi(t, x) |^2 = 1$.
The positional uncertainty of the wave packet at time $t=0$ is easily shown
to be equal to 
$\delta x(t)^2 = \langle x(t)^2 \rangle - \langle x(t) \rangle^2 = \delta x^2$
and thus constant in time (no dispersion of the wave packet).
Clearly, the wave packet describes an ultrarelativistic 
spinless particle whose wave function is centered at 
$\langle x(t) \rangle = c \, t$, but the tip of the wave
packet is forward displaced by a distance $\delta x$
relative to the center of the wave packet.
If the particle were classical, then its trajectory 
would be described by the expectation value
$\langle x(t) \rangle = c \, t$, i.e., the particle
would travel on the light cone.
However, quantum mechanically, 
it is possible for the tip of the wave packet 
to be ahead by a distance $\delta x$.
This observation alone, of course, by no means constitutes
a violation of causality, but it illustrates the fact
that one needs to be careful when analyzing the 
propagation of quantum mechanical wave packets.

The wave packet given in Eq.~\eqref{start} shows no 
dispersion over time, or, equivalently,
the uncertainty $[\delta x(t)]^2 = \delta x^2$ is
constant in time. Ultra-relativistic particles
traveling at exactly the speed of light are
described by wave packets which do not display dispersion,
because all momentum components travel at the speed of light
$c$, and the phase velocity $c$ is equal to the
group velocity $\partial E/\partial p = \partial( c \, p )/\partial p = c$.
One can understand the dispersion of quantum-mechanical
wave packets, in a somewhat colloquial, low-level
analogy,  by considering a herd of cows:
The faster cows will form the tip of the herd,
while the slower cows will stay behind.
The positional uncertainty $\delta x(t)^2$ of the 
herd grows with time.
The same phenomenon applies to an ``ultrarelativistic herd
of cows'' corresponding to the quantum mechanical wave packet
that describes a massive, ultra-relativistic Dirac particle.
Here, the dispersion relation is
$E = \sqrt{ p^2 + m^2 }$ (we assume $p = p_x$ to be 
the momentum in the $x$ direction),
and the phase velocity $E/p$ is not equal to the
group velocity $\partial E/\partial p = p/E < c$.
The dispersion relation $E = \sqrt{ p^2 + m^2 }$ 
describes particles traveling at speeds lower than
the speed of light (so-called tardyonic particles).
Recently, the tachyonic Dirac equation 
has been studied in detail~\cite{JeWu2012jpa}.
As is well known, the dispersion relation
for tachyonic particles reads as
$E = \sqrt{ p^2 - m^2 }$.
Again, the phase velocity $E/p$ is not equal to the
group velocity, yet the group velocity 
$\partial E/\partial p = p/E > c$ is superluminal.

When we detect ultrarelativistic particles on 
Earth of cosmic origin, a natural question to ask
concerns the spread of the quantum mechanically
``allowed'' arrival times, 
assuming that the wave packets describing particles, on their way 
through cosmos, propagated according to the free 
tardyonic and tachyonic Dirac equations.
Here, we thus engage in the interesting task to study the 
dispersion of wave packets solving the 
free tardyonic and tachyonic Dirac equations,
and to study the time evolution of the positional 
uncertainy under the free
tardyonic and tachyonic Dirac equations.

This paper is organized as follows:
In Sec.~\ref{sec2}, we study the bispinor solutions
(in the helicity basis) which constitute the 
basis of our considerations.
In Sec.~\ref{sec3}, we employ an expansion about the 
central momentum value $p_0$ of the wave packet 
in order to evaluate that positional 
uncertainty of the wave packet as a function
of time. The cosmic limit is discussed in Sec.~\ref{sec4}.
Conclusions are reserved for Sec.~\ref{sec5}.
Natural units with $\hbar = c = \epsilon_0 = 1$ 
are used everywhere in our calculations
unless stated otherwise.

\section{Bispinor Solutions}
\label{sec2}

\subsection{General Considerations}
\label{sec21}

In order to write the solutions for the 
tachyonic and tardyonic Dirac 
equations~\cite{ChHaKo1985,JeWu2012epjc,JeWu2012jpa,JeWu2013isrn,JeEtAl2014,NoJe2015tach,JeEh2016advhep,JeNaEh2017,SoNaJe2019,JeAd2022book},
\begin{subequations}
\label{dirac}
\begin{align}
\left( \ii \gamma^\mu \partial_\mu - \gamma^5 \, m \right) \, \Psi(x) = & \; 0 
\qquad
\mbox{(tachyonic)} \,,
\\
\left( \ii \gamma^\mu \partial_\mu - m \right) \, \psi(x) =& \; 0 
\qquad
\mbox{(tardyonic)} \,,
\end{align}
\end{subequations}
we resort to the the helicity basis 
adapted to the ultrarelativistic case~\cite{JeWu2012epjc}.
(Throughout this paper, quantities referring to 
tachyonic entities such as bispinor wave functions
are denoted by uppercase Greek and Latin symbols,
while quantities pertaining to tardyonic entities
are lowercase. Standard notation is used in Eq.~\eqref{dirac}
for the Dirac gamma matrices $\gamma^\mu$, the 
partial derivatives $\partial_\mu \equiv \partial/\partial x^\mu$,
the mass parameter $m$, and the space-time point $x = (t, \vec r)$.)

In order to meaningfully discuss the 
dispersion of an ultrarelativistic wave packet,
and study quantum propagation,
we need to construct normalizable states.
It is well known that momentum eigenstates are normalized to a
Dirac-$\delta$ in momentum space and their 
wave functions cannot be normalized to unity 
in coordinate space~\cite{JeAd2022book}.
This problem has been considered in various contexts before.
An example can be found in photon wave packets, 
which are interpreted as photon wave functions in Ref.~\cite{Mo2010}.
Specifically, in the discussion surrounding 
Eq.~(365) of Ref.~\cite{Mo2010},
Gaussian envelope factors are investigated.
Hermite--Gaussian modes (in momentum space) are discussed in Eq.~(122) of 
Ref.~\cite{SmRa2007} in order to discuss wave-packet 
quantization of photons, inspired by previous work~\cite{KoLi1966}
in this direction.

Here, we employ the Gaussian envelope function
\begin{equation}
f(p) = \frac{ ( 2 \pi )^{1/4} }{ \sqrt{ \delta p } } \, 
\exp\left( - \frac{(p-p_0)^2}{4 \delta p^2} \right) \,,
\end{equation}
which is normalized to unity
\begin{equation}
\int \frac{\dd p}{2 \pi} | f(p) |^2 = 1 \,,
\end{equation}
and has the property
\begin{equation}
\langle p^2 \rangle - \langle p \rangle^2 = \delta p^2 \,.
\end{equation}
The mean-square momentum uncertainty is equal to $\delta p^2$.

\subsection{Tachyonic Dirac spinors}
\label{sec22}

The tachyonic bispinor solutions have been studied by us in
Refs.~\cite{ChHaKo1985,JeWu2012epjc,JeWu2012jpa,JeWu2013isrn,%
NoJe2015tach,JeNaEh2017,SoNaJe2019}.
We recall the two-component helicity spinors as
\begin{equation}
a_+(\vec k) = \left( \begin{array}{c} 
\cos\left(\frac{\theta}{2}\right) \\[0.33ex]
\sin\left(\frac{\theta}{2}\right) \, \ee^{\ii \, \varphi} \\ 
\end{array} \right) \,,
\quad 
a_-(\vec k) = \left( \begin{array}{c}  
-\sin\left(\frac{\theta}{2}\right) \, \ee^{-\ii \, \varphi} \\[0.33ex]
\cos\left(\frac{\theta}{2}\right) \\
\end{array} \right) \,,
\end{equation}
where $\theta$ and $\varphi$ are the polar and azimuth angles
of the wave vector $\vec k$.
We start from the negative-helicity, positive-energy
solution given in Eqs.~(2.8) and (3.2b) of Ref.~\cite{JeWu2013isrn},
\begin{equation}
\Psi(t, \vec r) =
\left( \begin{array}{c}
\sqrt{\dfrac{|\vec p| - m}{2 \, |\vec p|}} \; a_-(\vec p) \\[2.33ex]
-\sqrt{\dfrac{|\vec p| + m}{2 \, |\vec p|}} \; a_-(\vec p) \\
\end{array} \right)\,
\exp\left( - \ii E \, t + \ii \vec p \cdot \vec r \right) \,,
\qquad E = \sqrt{ \vec p^{\,2} - m^2 } \,.
\end{equation}
For $\vec p = p_x \hate_x = p \hate_x$,
one has $\theta = 90^\circ = \pi/2$ and
$\varphi = 0$. Then,
\begin{equation}
\label{bispinor_tach}
\Psi(t, x, p) =
\frac{f(p)}{2} \, 
\left( \begin{array}{c}
- \sqrt{(p - m)/p }  \\[0.1133ex]
\sqrt{(p - m)/p }  \\[0.1133ex]
\sqrt{(p + m)/p }  \\[0.1133ex]
- \sqrt{(p + m)/p }  
\end{array} \right) \,
\exp\left( - \ii  \sqrt{ p^2 - m^2 }  \, t + \ii p \, x \right) \,,
\qquad
| \Psi(t, x, p) |^2 = f(p)^2 \,.
\end{equation}
A normalizable wave packet is thus obtained as
\begin{equation}
\Psi(t, x) =
\int \frac{\dd p}{2 \pi} 
\frac{f(p)}{2} \,
\left( \begin{array}{c}
- \sqrt{(p - m)/p }  \\[0.1133ex]
\sqrt{(p - m)/p }  \\[0.1133ex]
\sqrt{(p + m)/p }  \\[0.1133ex]
- \sqrt{(p + m)/p }
\end{array} \right) \,
\exp\left( - \ii \sqrt{ p^2 - m^2 } \, t + \ii p \, x \right) \,.
\end{equation}
It is normalized to
\begin{equation}
\int \dd x \, | \Psi(t, x) |^2 =
\int \dd x \, \Psi^+(t, x) \, \Psi(t, x) =
\int \frac{\dd p}{2 \pi} | f(p) |^2 = 1 \,.
\end{equation}
If one takes the ultrarelativistic limit ($p \gg m$) in the 
bispinor prefactor, one obtains a somewhat simpler form,
which illustrates the connection to the spinless case,
\begin{equation}
\label{approx_tach}
\Psi(t, x) \approx
\int \frac{\dd p}{2 \pi}
\frac{f(p)}{2} \,
{\underline u}
\exp\left( - \ii \sqrt{ p^2 - m^2 } \, t + \ii p \, x \right) \,,
\end{equation}
where ${\underline u} = (-1,1,1,-1)^{\rm T}$.
Because $f(p)$ is peaked in the region $p \approx p_0 \gg m$,
this integration region eventually dominates in all subsequent 
calculations.

\subsection{Tardyonic Dirac spinors}
\label{sec23}

We start from the negative-helicity, positive-energy
solution given in Eqs.~(2.8) and (2.10b) of Ref.~\cite{JeWu2013isrn},
\begin{equation}
\psi(t, \vec r) =
\left( \begin{array}{c}
\sqrt{\dfrac{E + m}{2 \, E}} \; a_-(\vec p) \\[2.33ex]
-\sqrt{\dfrac{E - m}{2 \, E}} \; a_-(\vec p)
\end{array} \right) \, 
\exp\left( - \ii E \, t + \ii \vec p \cdot \vec r \right) \,,
\qquad E = \sqrt{ \vec p^{\,2} + m^2 } \,.
\end{equation}
The solution describing a positive-energy,
negative-helicity particle is given as
\begin{equation}
\label{bispinor_tard}
\psi(t, x, p) =
\frac12 \,
\left( \begin{array}{c}
- \sqrt{(E + m)/E }  \\[0.1133ex]
\sqrt{(E + m)/E }  \\[0.1133ex]
\sqrt{(E - m)/E }  \\[0.1133ex]
- \sqrt{(E - m)/E }
\end{array} \right) \,
\exp\left( - \ii \sqrt{ p^2 + m^2 }  \, t + \ii p \, x \right) \,,
\qquad
| \psi(t, x, p) |^2 = 1 \,.
\end{equation}
A normalizable wave packet is obtained as
\begin{equation}
\psi(t, x) =
\int \frac{\dd p}{2 \pi}
\frac{f(p)}{2} \,
\left( \begin{array}{c}
- \sqrt{(E + m)/E }  \\[0.1133ex]
\sqrt{(E + m)/E }  \\[0.1133ex]
\sqrt{(E - m)/E }  \\[0.1133ex]
- \sqrt{(E - m)/E }
\end{array} \right) \,
\exp\left( - \ii E \, t + \ii p \, x \right) \,.
\end{equation}
It is normalized to
\begin{equation}
\int \dd x \, | \psi(t, x) |^2 =
\int \dd x \, \psi^+(t, x) \, \psi(t, x) =
\int \frac{\dd p}{2 \pi} | f(p) |^2 = 1 \,.
\end{equation}
If we take the ultrarelativistic limit ($p \gg m$) in the 
bispinor prefactor, then we can approximate the solution as
\begin{equation}
\label{approx_tard}
\psi(t, x) \approx
\int \frac{\dd p}{2 \pi}
\frac{f(p)}{2} \,
{\underline u}
\exp\left( - \ii \sqrt{ p^2 + m^2 } \, t + \ii p \, x \right) \,,
\end{equation}
where ${\underline u} = (-1,1,1,-1)^{\rm T}$.

\section{Standard Wave Packet}
\label{sec3}

\subsection{Tachyonic Case}
\label{sec31}

\begin{figure}[t!]%
\begin{center}
\begin{minipage}{0.9\linewidth}
\begin{center}\includegraphics[width=1.0\linewidth]{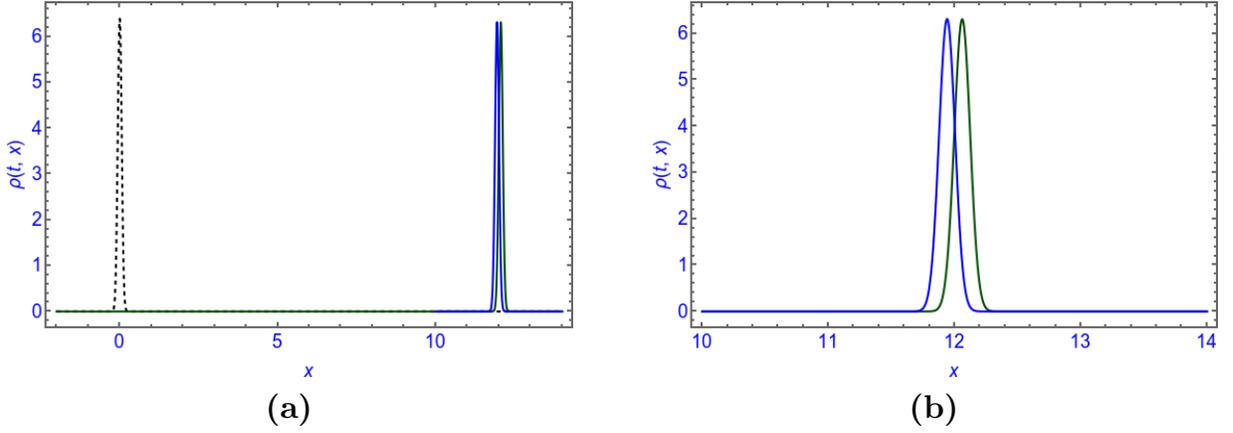}\end{center}%
\caption{\label{fig1} 
We illustrate the time-propagated tachyonic and tardyonic 
wave functions for the example case
$m= 10$, $p_0 = 100$, $\delta p = 8$, and $t_0 = 12$,
given in Eq.~\eqref{example}. 
The dashed curve in Fig.~(a) displays the initial 
density $\rho(t=0,x)$, while the blue curve 
shows the tardyonic density $\rho(t = t_0, x)$
and the dark green curve shows the 
tachyonic time-evolved function $R(t = t_0, x)$.
As demonstrated more clearly in the close-up in Fig.~(b),
the tachyonic wave has propagated a little faster 
in the positive $x$ direction as compared to the 
tardyonic wave. The positional uncertainty 
of the time-evolved tachyonic and tardyonic 
wave packets is almost the same, as is evident from 
Eqs.~\eqref{num_tach},~\eqref{num_tach2},~\eqref{num_tard},
\eqref{num_tard2} and~\eqref{result}. 
}
\end{minipage}
\end{center}
\end{figure}

We start from Eq.~\eqref{approx_tach}.
The tachyonic standard wave packet,
derived from Eq.~\eqref{approx_tach}, reads as
\begin{equation}
\label{standard_tach}
\Psi(t, x) = \frac{ ( 2 \pi )^{1/4} }{\delta p} \, \int \frac{\dd p}{2 \pi} 
\exp\left( -\ii \sqrt{ p^2 - m^2 } \, t + 
\ii p \, x - \frac{(p-p_0)^2}{4 \delta p^2} \right)
\end{equation}
It fulfills the (spinless) tachyonic wave equation,
\begin{equation}
\left( \frac{1}{c^2} \frac{\partial^2}{\partial t^2} - 
\frac{\partial^2}{\partial x^2} + m^2 \right) \, \Psi(t, x) = 0 \,,
\end{equation}
Its normalization is as follows,
\begin{equation}
\int \dd x \, | \Psi(t, x) |^2 = 1 \,,
\qquad
\langle X(t) \rangle = \int \dd x \, x \, | \Psi(t, x) |^2 \,,
\qquad
\langle X(t)^2 \rangle = \int \dd x \, x^2 \, | \Psi(t, x) |^2 \,.
\end{equation}
In the calculation of $\langle X(t) \rangle$,
one needs to consider three integrals,
namely, those over the two momentum parameters defining the 
wave functions, and one over the $x$ coordinate.
With advantages, one does the $x$ integral first,
with the help of the formula
\begin{equation}
\int \dd x \, x \, \exp\left( \ii (p - p') x\right) =
-\ii \frac{\partial}{\partial p} \delta(p - p')\,.
\end{equation}
One then  differentiates the integrand, 
and applies the Dirac-$\delta$ function,
reducing the problem to a one-dimensional $p$ integral
with an exponential weight factor.
In the last step, one does the remaining $p$ integral under the 
appropriate ultrarelativistic approximations.
The tachyonic expectations values read as follows,
\begin{align}
\langle [X(t)]^2 \rangle =& \; \frac{1}{4 \delta p^2} + t^2 +
\frac{m^2 \, t^2}{p_0^2} +
\frac{m^4 + 3 m^2 \delta p^2}{p_0^4} \, t^2
+ \frac{10 m^4 \delta p^2 + 15 m^2 \delta p^4 + m^6}{p_0^6} \, t^2
\nonumber\\
& \;
+ \frac{m^8 + 21 \, m^6 \, \delta p^2
+ 105 \, m^4 \, \delta p^4
+ 105 \, m^2 \, \delta p^6 }{p_0^8} t^2
+ \mathcal{O}(p_0^{-10}) \,,
\end{align}
for the position, and 
\begin{align}
[\langle X(t) \rangle]^2 =& \; t^2 +
\frac{m^2 \, t^2}{p_0^2} +
\frac{m^4 + 3 m^2 \delta p^2}{p_0^4} \, t^2
+ \frac{9 m^4 \delta p^2 + 15 m^2 \delta p^4 + m^6}{p_0^6} \, t^2
\nonumber\\
& \;
+ \frac{m^8 + 18 \, m^6 \, \delta p^2
+ \tfrac{177}{2} \, m^4 \, \delta p^4
+ 105 \, m^2 \, \delta p^6 }{p_0^8} t^2
+ \mathcal{O}(p_0^{-10})
\end{align}
for its square. The mean-square uncertainty of the 
position, as a function of time, is found as follows,
\begin{equation}
\label{deltaX2_tach}
\delta X(t)^2 =
\langle [X(t)^2] \rangle - [\langle X(t) \rangle]^2 = 
\frac{1}{4 \delta p^2} + \frac{m^4 \, \delta p^2 \, t^2}{p_0^6} +
\left( \frac{33 m^4 \, \delta p^4}{2 p_0^8} + 
\frac{3 m^6 \, \delta p^2}{p_0^8} \right) \, t^2 
+ \calO(p_0^{-10})
\end{equation}
At $t=0$, the Heisenberg uncertainty relation is fulfilled
in the minimal way, in the sense that 
$\delta x^2 \, \delta p^2 = 1/4$.
If one takes the bispinor prefactors from the 
spin-$1/2$ solution given in Eq.~\eqref{bispinor_tach}
into account, then some additional terms
are found for $t=0$,
\begin{multline}
\label{add_tach}
\langle [X(0)^2] \rangle_{s=1/2} - 
\langle [X(0)^2] \rangle_{s=0} =
\frac{m^2}{4 p_0^4}
+ \frac{1}{p_0^6} 
\left( \frac52 m^2 \delta p^2 + \frac14 m^4 \right)
\\
+ \frac{1}{p_0^8}
\left( \frac14 m^6 + \frac{21}{4} m^4 \delta p^2 +
\frac{105}{4} m^2 \delta p^4 \right) 
+ \calO(p_0^{-10}) \,.
\end{multline}
This approach leads to formulas for the 
time-dependent expectation values 
$\langle X(t)] \rangle$ and 
$\langle [X(t)^2] \rangle$, but does not 
discuss the time-dependent form of the wave function itself.
An approximate calculation of time propagated
wave function can, however, be accomplished as follows.
One starts from the representation
\begin{equation}
\Psi(t, x) = \frac{ ( 2 \pi )^{1/4} }{ \sqrt{\delta p} } \, \int \frac{\dd p}{2 \pi}
\exp\left( \ii \, \Phi(t, x, p) \right) \,,
\qquad
\Phi(t, x, p) =
- \sqrt{ p^2 - m^2 } \, t + p \, x + \ii \frac{(p-p_0)^2}{4 \delta p^2} \,.
\end{equation}
The dominant momentum region is around $p \approx p_0$.
One expands $\Phi(t, x,p)$ about $p= p_0$, 
up to second order in $(p=p_0)$, and integrates 
over the resulting Gaussian function in $p$,
after completing the square. The result for the 
density $| \Psi(t, x) |^2$ is finally found in a relatively 
compact form,
\begin{subequations}
\label{rho_tach}
\begin{align}
R(t, x) =& \; | \Psi(t, x) |^2 = 
\calN(t, x) \, \exp\left( -\calG(t, x) \right) \,,
\qquad
\calN(t, x) =
\sqrt{\frac{2}{\pi}} \, 
\frac{ (p_0^2 - m^2)^{3/2} \, \delta p }%
{ \sqrt{ ( p_0^2 - m^2 )^3 + 4 m^4 t^2 \delta p^4} } \,,
\\[0.1133ex]
\calG(t, x) =& \;
\frac{2 (p_0^2 - m^2)^2 \, \delta p^2}{ (p_0^2 - m^2)^3 + 4 m^4 t^2 \delta p^4 } 
\left[ p_0^2 (t^2 + x^2) - m^2 x^2 - 2 p_0 \sqrt{ p_0^2 - m^2 } \, t \, x \right]
\end{align}
\end{subequations}
Let us consider an example and temporarily restore SI units.
For illustration purposes, we consider
\begin{equation}
\label{example}
m = 10 \, {\rm u}\,, \qquad
p_0 = 100 {\rm u} \, c\,,  \qquad
\delta p = 8 {\rm u} \, c\,, \qquad
t_0 = 12 \frac{\hbar}{{\rm u} c^2} \,,
\end{equation}
where ${\rm u}$ is an arbitrarily chosen mass scale,
$c$ is the speed of light, and $\hbar$ is Planck's unit of action.
These parameters lead to a numerically and 
graphically convenient representation (see Fig.~\ref{fig1},
where in the figure, we set $\hbar = {\rm u} = c = 1$).
Numerically, one obtains the results
\begin{equation}
\label{num_tach}
\sqrt{ \langle [X(t_0)]^2 \rangle } = 12.061842 \frac{\hbar}{{\rm u} c}\,,
\qquad
             \langle X(t_0) \rangle = 12.061676 \frac{\hbar}{{\rm u} c} \,,
\qquad
\delta \langle [X(t_0)]^2 \rangle   = 0.0040125 \left( \frac{\hbar}{{\rm u} c} \right)^2 \,.
\end{equation}
The first terms listed in Eq.~\eqref{deltaX2_tach} add up to
\begin{equation}
\label{num_tach2}
\frac{\hbar^2}{4 \delta p^2} + \frac{(m c)^4 \, \delta p^2 \, (c t)^2}{p_0^6} +
\left( \frac{33 (m c)^4 \, \delta p^4}{2 p_0^8} + 
\frac{3 (m c)^6 \, \delta p^2}{p_0^8} \right) \, (c t)^2 
= 0.0040109 \, \left( \frac{\hbar}{{\rm u} c} \right)^2 \,,
\end{equation}
leading to very good agreement with the analytic result~\eqref{deltaX2_tach}.
Note that 
the initial mean-square positional uncertainty is
$\delta \langle [X(t = 0)]^2 \rangle   = 0.00390625 (\hbar/({\rm u} c))^2$,
which is manifestly different from the result given in 
Eq.~\eqref{num_tach}. We now switch back to natural units.

%
%

\subsection{Tardyonic Case}
\label{sec32}

We start from Eq.~\eqref{approx_tard} and
define a tardyonic standard wave packet,
\begin{equation}
\label{standard_tard}
\psi(t, x) = \frac{\sqrt{2 \pi}}{\delta p} \, \int \frac{\dd p}{2 \pi} 
\exp\left( -\ii \sqrt{ p^2 + m^2 } \, t + 
\ii p \, x - \frac{(p-p_0)^2}{4 \delta p^2} \right) \,.
\end{equation}
It solves the tardyonic (subluminal) wave equation,
\begin{equation}
\left( \frac{1}{c^2} \frac{\partial^2}{\partial t^2} -
\frac{\partial^2}{\partial x^2} - m^2 \right) \, \Psi(t, x) = 0 \,.
\end{equation}
The normalization is as follows,
\begin{equation}
\int \dd x \, | \Psi(t, x) |^2 = 1 \,,
\qquad
\langle x(t) \rangle = \int \dd x \, x \, | \Psi(t, x) |^2 \,,
\qquad
\langle x(t)^2 \rangle = \int \dd x \, x^2 \, | \Psi(t, x) |^2 \,.
\end{equation}
The tardyonic expectation value of the mean-square position is
\begin{align}
\langle [x(t)]^2 \rangle =& \; \frac{1}{4 \delta p^2} + t^2 -
\frac{m^2 \, t^2}{p_0^2} +
\frac{m^4 - 3 m^2 \delta p^2}{p_0^4} \, t^2 
+ \frac{10 m^4 \delta p^2 - 15 m^2 \delta p^4 - m^6}{p_0^6} \, t^2
\nonumber\\
& \;
+ \frac{m^8 - 21 \, m^6 \, \delta p^2 
+ 105 \, m^4 \, \delta p^4
- 105 \, m^2 \, \delta p^6 }{p_0^8} t^2 
+ \mathcal{O}(p_0^{-10}) \,.
\end{align}
The square of the expected position is
\begin{align}
[\langle x(t) \rangle]^2 =& \; t^2 -
\frac{m^2 \, t^2}{p_0^2} +
\frac{m^4 - 3 m^2 \delta p^2}{p_0^4} \, t^2
+ \frac{9 m^4 \delta p^2 - 15 m^2 \delta p^4 - m^6}{p_0^6} \, t^2
\nonumber\\
& \;
+ \frac{m^8 - 18 \, m^6 \, \delta p^2
+ \tfrac{177}{2} \, m^4 \, \delta p^4
- 105 \, m^2 \, \delta p^6 }{p_0^8} t^2
+ \mathcal{O}(p_0^{-10})
\end{align}
The mean-square coordinate uncertainty is 
thus
\begin{equation}
\label{deltax2_tard}
\delta x(t)^2 = 
\langle [x(t)^2] \rangle - [\langle x(t) \rangle]^2 =
\frac{1}{4 \delta p^2} +
\frac{m^4 \, \delta p^2 \, t^2}{p_0^6} +
\left( 
\frac{33 m^4 \, \delta p^4}{2 p_0^8} -
\frac{3 m^6 \, \delta p^2}{p_0^8} 
\right) \, t^2 + \mathcal{O}(p_0^{-10}) \,,
\end{equation}
which is seen to be equivalent to the 
result given in Eq.~\eqref{deltaX2_tach} 
upon the replacement $m \to \ii \, m$.
One finds, for the solution in Eq.~\eqref{bispinor_tard},
additional terms at the initial time $t = 0$,
\begin{multline}
\label{add_tard}
\langle [x(0)^2] \rangle_{s=1/2} - 
\langle [x(0)^2] \rangle_{s=0} =
\frac{m^2}{4 p_0^4} 
+ \frac{1}{p_0^6} 
\left( \frac52 m^2 \delta p^2 - \frac12 m^4 \right)
\\
+ \frac{1}{p_0^8} 
\left( \frac34 m^6 - \frac{21}{2} m^4 \delta p^2 +
\frac{105}{4} m^2 \delta p^4 \right)
+ \calO(p_0^{-10}) \,.
\end{multline}
For the approximate calculation of time propagated
wave function, one employs the same steps that 
lead to Eq.~\eqref{rho_tach} in the tachyonic case.
One writes the wave function as
\begin{equation}
\psi(t, x) = \frac{\sqrt{2 \pi}}{\delta p} \, \int \frac{\dd p}{2 \pi}
\exp\left( \ii \phi(t, x, p) \right) \,,
\qquad
\phi(t, x, p) =
- \sqrt{ p^2 + m^2 } \, t + p \, x + \ii \frac{(p-p_0)^2}{4 \delta p^2} \,.
\end{equation}
Then, one expands $\phi(t, x,p)$ about $p= 0$, 
up to second order in $(p=p_0)$, and integrates 
over the resulting Gaussian function in $p$,
after completing the square. The result is
\begin{subequations}
\label{rho_tard}
\begin{align}
\rho(t, x) =& \; | \psi(t, x) |^2 =
n(t, x) \, \exp\left( -g(t, x) \right) \,,
\qquad
n(t, x) =
\sqrt{\frac{2}{\pi}} \,
\frac{ (p_0^2 + m^2)^{3/2} \, \delta p }%
{ \sqrt{ ( p_0^2 + m^2 )^3 + 4 m^4 t^2 \delta p^4} } \,,
\\[0.1133ex]
g(t, x) =& \;
\frac{2 (p_0^2 + m^2)^2 \, \delta p^2}{ (p_0^2 + m^2)^3 + 4 m^4 t^2 \delta p^4 }
\left[ p_0^2 (t^2 + x^2) + m^2 x^2 - 2 p_0 \sqrt{ p_0^2 + m^2 } t \, x \right] \,.
\end{align}
\end{subequations}
We consider the same example as in Eq.~\eqref{example},
$m= 10 \, {\rm u}$, $p_0 = 100 \, {\rm u} \, c$, 
$\delta p = 8 \, {\rm u} \, c$, and $t_0 = 12 \hbar/({\rm u} c^2)$,
but for the tardyonic case,
and temporarily switch to SI units (again).
Numerically, one obtains the results
\begin{equation}
\label{num_tard}
\sqrt{ \langle [X(t_0)]^2 \rangle } = 11.939453 \frac{\hbar}{{\rm u} c} \,,
\qquad
             \langle X(t_0) \rangle = 11.939286 \frac{\hbar}{{\rm u} c} \,,
\qquad
\delta \langle [X(t_0)]^2 \rangle   = 0.0040058 \left( \frac{\hbar}{{\rm u} c} \right)^2 \,.
\end{equation}
The first terms listed in Eq.~\eqref{deltax2_tard} add up to
\begin{equation}
\label{num_tard2}
\frac{\hbar^2}{4 \delta p^2} + \frac{(m c)^4 \, \delta p^2 \, (c t)^2}{p_0^6} +
\left( \frac{33 \, (m c)^4 \, \delta p^4}{2 p_0^8} +
\frac{3 \, (m c)^6 \, \delta p^2}{p_0^8} \right) \, (c t)^2
= 0.0040054 \left( \frac{\hbar}{{\rm u} c} \right)^2 \,.
\end{equation}
This result is in very good agreement with the analytic 
result~\eqref{deltax2_tard}.
A plot, which illustrates the difference between
tachyonic and tardyonic propagation, is found in Fig.~\ref{fig1}
(where we set $\hbar = {\rm u} = c = 1$ in the plot).
From now on, we again switch to natural units.

\section{Cosmic Limit}
\label{sec4}

Up to this point, we have
assumed that $p_0$ is the largest variable in the problem.
The cosmic limit is obtained when one considers
large propagation times and distances.
It is appropriate to scale $t \to \lambda \, t$ and 
$x \to \lambda \, x$ and to keep only the leading-order 
terms in $\lambda$.
A careful investigation of the expressions 
for $R(t,x)$ and $\rho(t,x)$
is sufficient.
One can, in fact, integrate over $x$ and $x^2$
with the densities given in Eqs.~\eqref{rho_tach} and~\eqref{rho_tard},
and calculate the
standard deviation of the position expectation value.
The limit of large $t$, 
small $m$, and large $p_0$ is obtained as
\begin{equation}
\label{cosmic}
\delta X(t)^2 \approx
\delta x(t)^2 \approx
\frac{m^4 \, \delta p^2 \, t^2}{p_0^6}  \,,
\qquad 
t \to \infty \,.
\qquad 
\frac{\delta p}{p_0} \ll 1 \,,
\qquad 
\frac{m}{p_0} \ll 1 \,.
\end{equation}
In order to confirm and ramify the result,
one observes that the limit of large $t$, and $m \ll \delta p \ll p_0$
means that the $1/p_0^8$ terms in 
Eqs.~\eqref{deltaX2_tach} and~\eqref{deltax2_tard}
are suppressed in comparison to the $1/p_0^6$ term,
which is listed in Eq.~\eqref{cosmic}.
Finally, since we are investigating the 
ultrarelativistic limit,
it is useful to convert the positional
uncertainty into a detection time uncertainty 
acquired for the detection of the 
ultrarelativistic particle coming in from
the cosmos, and convert the result to SI mksA units.
We choose as the cosmic travel time an interval
of 168,000 light years,
which is the distance to the Large Magellanic Could,
where the supernova 1987A originated~\cite{DaEtAl1987}.
One finds
\begin{equation}
\label{result}
\left. \frac{\delta X(t)}{c} \right|_{t = 168,000 \, {\rm yr}} \approx
\left. \frac{\delta x(t)}{c} \right|_{t = 168,000 \, {\rm yr}} \approx
5.298 \times 10^{-6} \,
\frac{\delta \xi}{\xi} \; \left( \frac{\chi}{\xi} \right)^2 \, {\rm s} \,,
\end{equation}
where ``s'' of course is the symbol for the unit ``second'',
$\delta \xi$ is the momentum spread in ${\mathrm{GeV}}/c$,
$\xi$ is equal to the central momentum $p_0$ in ${\mathrm{GeV}}/c$,
and $\chi$ is the mass of the particle, measured in 
${\mathrm{eV}}/c^2$.
It means that, if the particle wave function is centered
about a well-defined ultrarelativistic mean momentum
$p_0 \gg m$ (i.e., $\delta \xi / \xi \ll 1$ and
$\chi/\xi \ll 1$),
then the detection time uncertainty amounts to
less than a microsecond even for cosmic travel 
over appreciable distances
(here, as an example, the distance to the Large Magellanic Could).
The result~\eqref{result} applies equally to tachyons as well as tardyons.

%
%
\section{Conclusions}
\label{sec5}

In this paper, we have investigated the propagation
of ultrarelativistic tachyonic and tardyonic wave packets
on short as well as cosmic time and distance scales.
In Sec.~\ref{sec2}, we have studied the 
positive-energy bispinor solutions of left helicity of the 
tachyonic and tardyonic Dirac equations, 
which describe propagation in the positive $x$ direction
of an ultra-relativistic spin-$1/2$ particle.
We use these as exemplary states in order to 
study the problem at hand.
Our choice is inspired by the fact that 
neutrinos, which remain a possible candidate 
for tachyons, typically occur in left-helicity states.
For an up-to-date summary of possible empirical 
indications that some neutrinos are tachyons,
see Ref.~\cite{Eh2022}.
We find that, in the ultrarelativistic limit,
the spin-$1/2$ solution, in the helicity basis,
reduces to the spinless solution, multiplied
by a constant four-components bispinor
[see Eqs.~\eqref{approx_tach},~\eqref{approx_tard}
and~\eqref{standard_tach} as well as~\eqref{standard_tard}].

A normalizable, localizable
wave packet of Gaussian shape is used as an initial
condition for the time propagation (see Sec.~\ref{sec3}.
We find that the mean-square positional
uncertainty of the wave packet increases quadratically
with time, for both tachyonic as well as 
tardyonic particles.
Our results (for the spinless case) 
are summarized in Eqs.~\eqref{deltaX2_tach}
and~\eqref{deltax2_tard}.
For spin-$1/2$ particles, 
the result is (somewhat surprisingly) almost the same,
but the initial value (at $t=0$) for the mean-square 
positional uncertainty receives a modification according to 
Eqs.~\eqref{add_tach} and~\eqref{add_tard}.

In the cosmic, ultrarelativistic limit ($m \ll p_0$),
the time $t$ is the dominant variable,
and we assume that the momentum uncertainty $\delta p$
is also much smaller than $p_0$.
In this limit, the result for the time evolution
of the positional uncertainy reduces to a single
term, given in Eq.~\eqref{cosmic}.
This result is proportional to $m^4$, where $m$ is the 
mass term, and hence invariant under the
replacement $m \to \ii \, m$.
A numerical evaluation of the result given in 
Eq.~\eqref{cosmic} reveals that,
under reasonable assumptions about the 
momentum spread of ultrarelativistic particles,
the dispersion of ultrarelativistic wave packets is sufficiently
small, even on cosmic times scales,
that it remains possible to associate 
the generation time of the particle with the 
detection time, up to an uncertainty which 
increases by no more than a microsecond 
per light year [see Eq.~\eqref{result}].
One of the consequences of this result is that 
the ``early'' arrival of the neutrino 
burst under the Mont Blanc, recorded in coincidence
with the 1987A supernova, cannot be explained by 
the positional uncertainty of the propagated
ultrarelativistic wave packet on its way from the 
Large Magellanic Cloud. 
[Other possible explanations (e.g., Ref.~\cite{Eh2018})
have been discussed in the literature.]
Our result~\eqref{result} is generally applicable
for tachyons and tardyons.

Finally, let us explore if we can understand the 
result~\eqref{result} intuitively,
with respect to the ``herd of cows analogy''
made in Sec.~\ref{sec1}.
In a wave packet composed of a tardyonic herd,
in view of the classical dispersion relation 
$E = m/\sqrt{1 - v^2}$ (with $v < 1$),
the faster ``cows'' are the ones with more energy.
In a tachyonic herd, 
in view of the classical dispersion relation 
$E = m/\sqrt{v^2 - 1}$ (with $v > 1$),
the faster ``cows'' are ones are the ones with less energy.
in both cases, the herd is composed of faster and slower ones,
and the wave packet spreads.
This consideration qualitatively explains the
universal character of the
result~\eqref{result} for tachyons and tardyons.

\acknowledgments

The authors acknowledge helpful conversations
with Jos\'{e} Crespo L\'{o}pez--Urrutia on general
aspects regarding tachyons. Support from the
National Science Foundation (Grant PHY--2110294)
also is gratefully acknowledged.

\end{document}